\documentstyle[11pt,aas2pp4]{article}
\def\beq{\begin{equation}}
\def\eeq{\end{equation}}
\def\bey{\begin{eqnarray}}
\def\eey{\end{eqnarray}}
\def\kms{\mbox{\rm \,km\,s}^{-1}}
\input epsf

\begin{document}

\title
	{The survival of the Sgr dwarf galaxy and the flatness of
the rotation curve of the Galaxy}
\author
	{HongSheng Zhao
	\\Sterrewacht Leiden, 
Niels Bohrweg 2, 2333 CA, Leiden, The Netherlands (hsz@strw.LeidenUniv.nl)}
\date{Accepted ........      Received .......;      in original form .......}
\label{firstpage}

\begin{abstract}
How has the ``fluffy'' core of the Sgr dwarf galaxy survived multiple
strong shocks from the tidal force of the Galactic halo and disc since
the formation of the core a Hubble time ago?  A scenario that Sgr was
deflected to its current orbit by the Magellanic Clouds after a
rendezvous on the north Galactic pole $2-3$ Gyrs ago is examined.  It
is shown that the conditions of the collision fix both the sense of
circulation of Sgr and the LMC around the Galaxy and the slope of the
Galactic rotation curve.  The model argues that the two orthogonal
polar circles traced by a dozen or so Galactic halo dwarf galaxies and
globular clusters (LMC-SMC-Magellanic Stream-Draco-Ursa Minor along $l
\approx 270^o$ and M54-Ter 7-Ter 8-Arp 2-NGC 2419-Pal 15 along $l
\approx 0^o$) are streams of tidal relics from two ancient galaxies
which was captured on two intersecting polar rosette orbits by the
Galaxy.  Our results favor the interpretation of microlensing towards
the LMC being due to source or lens stars in tidal features of the
Magellanic Clouds.  We discuss direct and indirect observations to
test the collision scenario.
\end{abstract}


\section{Introduction}

The recently discovered dwarf galaxy at about $25$ kpc from the Sun in
the direction of the Sagittarius constellation (Ibata, Gilmore \&
Irwin 1994) is the closest galaxy known to us.  It is traced by two
long trailing/leading tails on the sky (together more than $8^o\times
22^o$ in solid angle) with most of its stars still clustered around a
low density luminous core (roughly $0.001 L_\odot{\rm pc}^{-3}$ with
semi-axes $1:1:3$ kpc).  It is puzzling why this fluffy core of the
dwarf galaxy has not been fully ``digested'' by the Galaxy, in the
sense that stars have not fully dispersed out of the core despite the
severe shocks at pericentric passage from the tidal force of the
Galactic halo (about $10-100$ times stronger than that experienced by
satellites in the outer halo, the Magellanic Clouds and the Fornax
dwarf galaxy included) and shocks when crossing the disc of the
Galaxy.  The best fit to Sgr's morphology, radial velocity (Ibata,
Gilmore \& Irwin 1995) and proper motion (Ibata, Wyse, Gilmore \&
Suntzeff 1997) yields an orbit with a pericenter-to-pericenter period
of about $0.8$ Gyr and a peri and apo-center at about 10 and 50 kpc
respectively (Vel\'azquez \& White 1995).  Simulations show that if a
typical Galactic dwarf galaxy (such as Fornax) were replaced on Sgr's
orbit, it would dissolve in no more than two peri-centric passages by
the strong peri-centric tidal shock of the Galaxy near 10 kpc
(Vel\'azquez \& White 1995; Johnston, Spergel \& Hernquist 1995;
Johnston, Hernquist \& Bolte 1996; Edelsohn \& Elmegreen 1997).  This
apparently contradicts the observation that the dominant stellar
population in the core is older than 10 Gyrs (Mateo et al. 1995,
Fahlman et al. 1996), implying that Sgr has survived $10-20$
peri-centric tidal shocks of the Galaxy.

To circumvent this dilemma we need to abandon either or both of the
following hidden assumptions: (i) the light distribution of Sgr traces
its mass, and (ii) Sgr has always been on the same low-pericentric
orbit in a rigid Galactic potential for the past 5 to 10 Gyrs.  Ibata,
Wyse, Gilmore \& Suntzeff (1997) postulate a dense dark halo of Sgr
surrounding the luminous part to hold the system together; they
require Sgr's mass density to be uniform inside about 3 kpc of its
core with a value ($\sim 0.03 M_\odot{\rm pc}^{-3}$) several times the
mean Galactic halo density inside $10$ kpc ($0.013M_\odot{\rm
pc}^{-3}$).  An inspection of Sgr's rosette-like orbit in relation to
that of the Magellanic Clouds (MCs) offers a completely different line
of thought.  They are on nearly orthogonal planes intersecting along
the poles with their Galactocentric radii overlapping at about 50 kpc.
So an encounter at the north or south pole some time in the past or
future is quite inevitable.  A recent preprint by Ibata \& Lewis
(1998), shortly after the completion of the work reported in this {\it
Letter}, also remarked on a small chance of an interaction after
noticing in their simulations a weak perturbation to Sgr's orbit when
they turned on the moving gravitational field of the massive MCs.
Unfortunately the effect was in the end neglected on grounds of low
probability without thoroughly exploring the parameter space (of
satellite velocities and the Galactic potential) and the important
consequences of a rare strong interaction.  So same as Ibata et
al. (1997) they were thus left with no alternative but to conclude a
massive dark halo of Sgr to be the only explanation for Sgr's presence
on a low-pericentric orbit after a Hubble time.

In this {\it Letter} we examine the encounter scenario, as illustrated
in Fig. 1, where Sgr has been pulled back from an originally high
angular momentum/energy orbit to the present low angular
momentum/energy orbit by the massive MCs.  A recent encounter would
have the advantage to allow Sgr to spend most of its lifetime on a
``safe'' orbit with a pericenter (say, $20$ kpc) too high to be
harassed by the sharply declining tidal force of the Galaxy (e.g.,
Kroupa 1997, Oh, Lin \& Aarseth 1995); in
a halo with an $r^{-2}$ density profile the pericentric shock would
drop by a factor of $4$ from $10$ kpc to $20$ kpc.  

\onecolumn
\begin{figure}
\epsfysize=15cm \centerline{\epsfbox{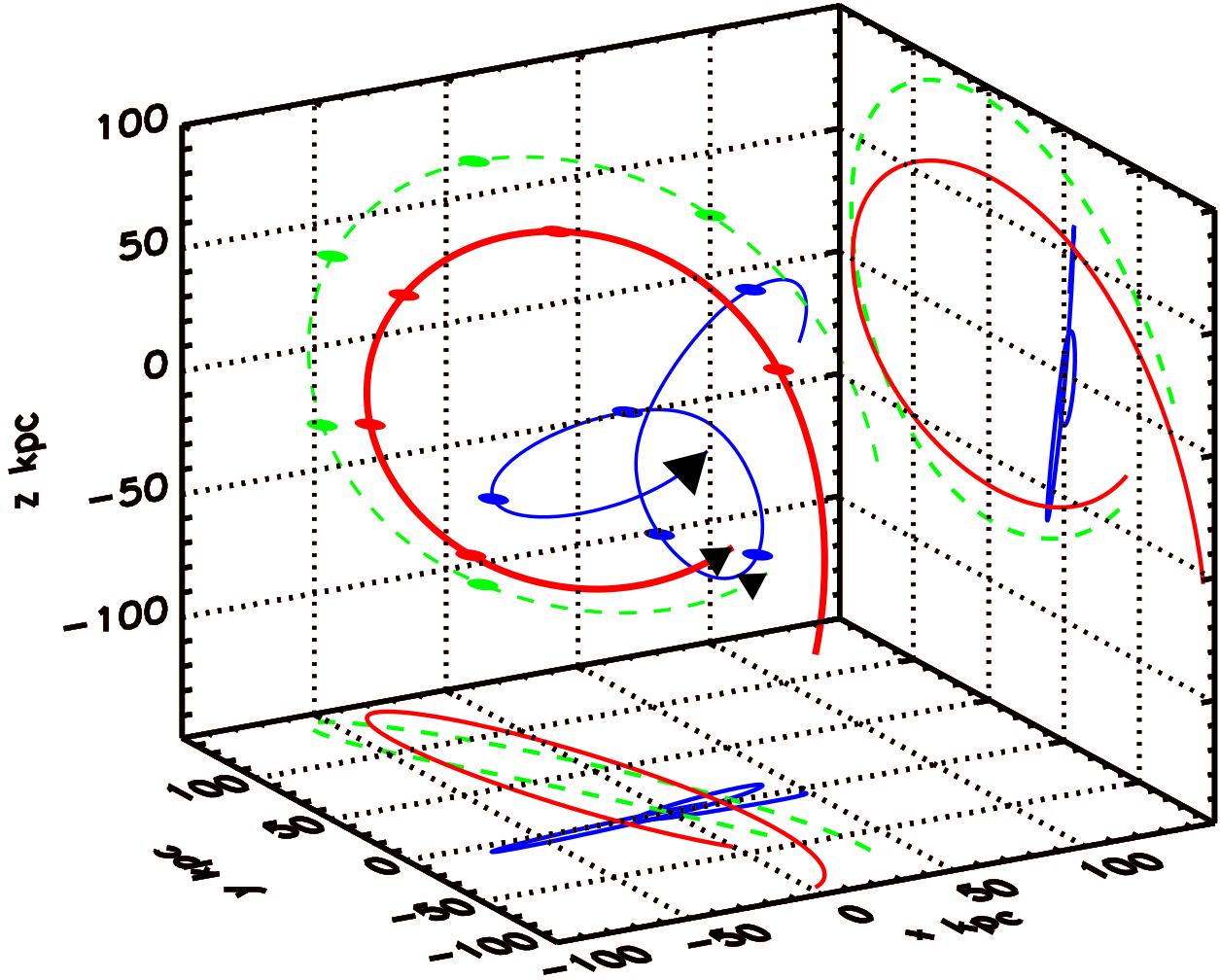}}
\caption{
A 3D view and $x-y$, $y-z$ projections of the orbits of the LMC/SMC 
(thick red/dashed green lines near the $x=0$ plane) and 
Sgr (thin blue curve near the $y=0$ plane);
the Sun is be at $(x,y,z)=(-8,0,0)$ kpc.
The three systems are integrated backward for $3$ Gyrs (with ellipses
marking steps of $0.5$ Gyr) from the present epoch 
(with velocity vectors marked by arrows) inside 
the Galactic potential.\label{sgrmc3d}
}
\end{figure}
\twocolumn

Various interesting aspects of this scenario will be discussed at the
end.  But the aim of this {\it Letter} is to report an independent
constraint on the rotation curve of the Galaxy as imposed purely by
{\it timing the collision}.  The essence is the following.  The random
chance for the LMC and Sgr to meet each other is obviously low, about
1\% for a 10 kpc closest approach in the past 3 Gyrs for a general set
of Galactic potentials and initial conditions of the satellites.  So
the same argument could be inverted: once we accept the deflection by
the MCs as a plausible way out of Sgr's dilemma a stringent set of
conditions on the potential of the halo and the proper motions of the
satellites must follow.

\section{``Measure'' the potential of the MW and proper motions of the LMC and Sgr by timing}

Consider the following timing argument for when and where the
collision happens.  The collision happens most likely when Sgr was
near its apocenter and the LMC again near its pericenter which means
that Sgr was $\left(n+{1 \over 2}\right)$ 
epicycles back and the LMC $k$ epicycles back,
where $n$ and $k$ are integers.  The angles which the LMC and Sgr have
rotated away from the site of the collision are related to the
epicycles by
$
\beta \approx {360^o k \over \theta_{lmc}} \approx {360^o (n+0.5)
\over  \theta_{sgr}},
$
where $1<\beta<2$ is the ratio of the period of one rotation
around the Galaxy with that of one radial epicycle; it is essentially
a constant for all orbits in a nearly power-law potential (Binney \&
Tremaine 1987; Johnston 1997).  Since 
Sgr ($l=5.6^o$,$b=-14.1^o$) and the LMC ($280^o$,$-33^o$)
are presently $250^o$ and $240^o$ 
from the north Galactic pole, to meet at the poles requires 
$\left[{\pm  \theta_{lmc} - 240^o \over 360^o},
{\pm  \theta_{sgr}-250^o \over 360^o}\right] $
to approximate to a pair of integers
if the collision was on the north pole or half integers if on
the south pole, where the plus sign corresponds to Sgr and the LMC
moving towards the plane and the minus sign, moving away.  
In solving the above equations we allow
$\pm 20^o$ angular offset from the poles and 
$\pm 45^o$ phase offset (equivalent of ${1 \over 8}$ of an epicycle)
from the pericenters or apocenters at the time of collision.
This roughly puts both satellites at around 10 kpc of each other, and
at about $40-60$ kpc from the Galactic center.  
The dynamical friction
with the halo can offset the sky position of the massive LMC at the
time of the encounter, but the amount $\Delta \theta \sim  {M_{lmc} k
\over M_{mw}(r<50 {\rm kpc})} \sim {(1.5\pm
1.0)\times 10^{10}M_\odot \over  5\times 10^{11}M_\odot}k
\sim (0.03 \pm 0.02) \sim 3^o$ 
is neglegibly small compared to the allowed error of $20^o$; in addition
recent data by Kunkel et al. 1997 favors a small mass for the LMC.
With a similar argument any slight
flattening in the Galactic potential can be neglected; squashing the
Galactic potential to an axis ratio of $q=0.9$ would change the orbit
by a tolerable amount of $\Delta \theta \sim (1-q) \sim 0.1 \sim 6^o$.

It is easy to show that as far as recent collisions are concerned (say
$n<3$), the only possible solution is that $n=2$, $k=1$, $
\theta_{lmc} = (240^o \pm 20^o)$, $ \theta_{sgr} = (610^o \pm
20^o)$.  This means that the collision happened on the 
{\it north Galactic pole}, the
LMC is presently leading the Magellanic Stream, and the Sgr is moving
{\it towards} the Galactic plane (exactly in the sense as the observed
proper motion of Sgr by Ibata et al. 1997).  The timing argument also
predicts that the epicycle period of the LMC is about ${\theta_{sgr}
\over \theta_{lmc}} = (2.54 \pm 0.23)$ times that of Sgr, which
matches very well with $2$ Gyrs and $0.8$ Gyrs epicycle periods for
the LMC (e.g., Lin, Jones, \& Klemola 1995; Moore \& Davis 1994;
Gardiner, Sawa \& Fujimoto 1994;
Murai \& Fujimoto 1980; 
Lin \& Lynden-Bell 1982 and references therein) and Sgr
(e.g. Vel\'azquez \& White 1995, Ibata et al. 1997)
respectively from previous models.  Now Sgr has circulated around the
Galaxy $(610^o\pm 20^o)$ from the start of the collision and in the
meantime advanced $2.5\pm 0.125$ epicycles, equivalent to a phase
angle $(900^o \pm 45^o)$.  Thus an estimate can also be made of
$\beta$, the ratio of the rotation period to the epicycle period in
the Galactic potential: $ \beta = {900^o\pm 45^o
\over 610^o\pm 20^o} = (1.475 \pm 0.088)$.  Combined with a similar
estimate from the LMC's position, we have $\beta= (1.48 \pm 0.08)$,
close to the value ($\sqrt{2}$) for a logarithmic potential.  This
provides a fully {\it independent} argument for a dark halo of the Milky
Way at intermediate radius (10-100 kpc)
where the constraints from traditional data sets are weak.  

An indirect ``measure'' of the velocities of the Magellanic Clouds and
Sgr can also be made with similar analytical arguments.  The velocity of a
satellite at radius $r$ is related to 
the characteristic size ($R$) of its orbit simply by
$V=\sqrt{2[\Phi(R)-\Phi(r)]}=V_{\!c,mw}\sqrt{2 \ln(R/r)}$ 
in a logarithmic potential $\Phi(r)=V_{\!c,mw}^2 \ln (r)$;
for an exactly radial orbit $R$ is the apocenter radius.  Thus 
a close encounter of the LMC and Sgr requires
\bey
{V_{lmc} \over V_{\!c,mw}} &=& \sqrt{2 \ln\left({R_{lmc} \over r_{lmc}}\right)} = (1.37 \pm 0.27),\\\nonumber
{V_{sgr}\over V_{\!c,mw}} &=& \sqrt{2 \ln\left({R_{sgr} \over r_{sgr}}\right)}
= (1.51 \pm 0.15).
\eey
In the above estimation we have adopted
$r_{sgr}=(16 \pm 1)$ kpc and $r_{lmc}=(50 \pm 1)$ kpc for
the present radii of the two satellites.  For Sgr to reach
the LMC $R_{sgr}$ should equal $r_{lmc}$ with a 10 kpc
uncertainty, thus $R_{sgr} = (50 \pm 10)$ kpc.  
To estimate $R_{lmc}$, we 
note that the orbital period is
nearly proportional to the orbital size $a$ in a logarithmic potential
(Johnston 1997), thus $R_{lmc} = (2.54 \pm 0.23)R_{sgr}= (127 \pm 40)$ kpc.  

Two interesting results follow from the above condition.  First the
transverse velocity of Sgr can be estimated from that of the LMC (
Jones, Klemola \& Lin 1994 and HIPPARCOS measurements from Kroupa \&
Bastian 1997) from the velocity ratios after taking into account of
radial velocities.  The result $V_{t,sgr}=(237 \pm 60) \kms$ is
consistent with $V_{t,sgr}=(250 \pm 90) \kms$ from direct measurement
of Sgr's latitudinal proper motion with respect to the Galactic bulge
(Ibata et al. 1997).

Second the observed space velocities of Sgr and the LMC translate to a
circular rotation speed of the MW 
$V_{\!c,mw} = (200 \pm 53)\kms$ and $(194\pm 45)\kms$ respectively.
Approximating the rotation curve as a power law
with a slope ${d\log V_{\!c,mw} \over d \log r} = {\beta^2 \over 2}
-1 = 0.09 \pm 0.08$ and an amplitude normalized to 
$(197 \pm 34)\kms$ at 50 kpc, we find that 
the collision requires the dynamical mass of the Milky Way to increase as
\beq\label{logr-logm} 
\log \left[{M_{mw}(<r) \over 4.55 \times
10^{11}M_\odot}\right] = (1.2 \pm 0.3) \log \left[{r \over 50{\rm
kpc}}\right] \pm 0.3
\eeq 
in the radii from $10$ kpc to $127 \pm 40$ kpc 
as spanned by the orbits of Sgr and the LMC, where $2\sigma$ error bars
are used.

\section{Implications of the collision}

\begin{figure}
\epsfysize=15cm \centerline{\epsfbox{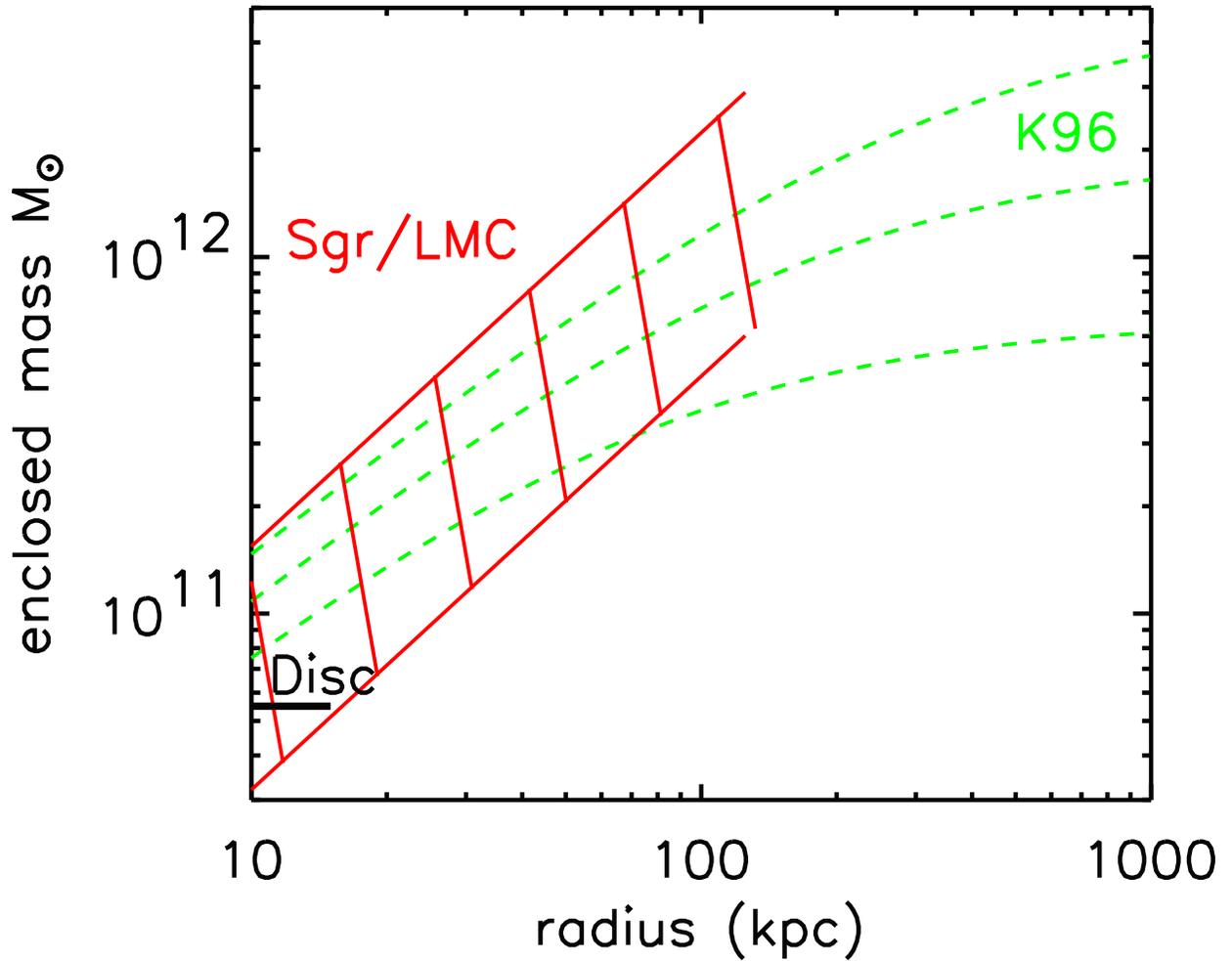}}
\caption{
compares radial distributions of the dynamical mass of the Galaxy 
derived from timing the collision of Sgr and the LMC (shaded region
with $2\sigma$ limits) and from Kochanek (1996), which synthesizes well-known
constraints from several data sets
(dashed lines for the median and $2\sigma$ limits).  Our model predicts
a dynamical mass of the Galaxy 10-40 times greater than 
the mass of a standard disc (heavy solid line segment).
\label{kochanek}
}
\end{figure}

In summary Sgr is proposed to have spent most of its lifetime on a
high-pericentric orbit based on the theoretical consideration that its
``fluffy'' core cannot sustain the repeated strong tidal shocks of a
low-pericentric orbit for 10 Gyrs.  A natural mechanism to bring Sgr
down to its present low orbit is a recent deflection by the passing
LMC.  By timing such a collision we get as a by-product an indirect
``measure'' of the $\log r - \log M$ relation of the Galaxy, both the
mean slope and zero point (cf. eq.~\ref{logr-logm}).
Figure~\ref{kochanek} compares our results with a previous
comprehensive analysis by Kochanek (1996).  Our analysis strengthens
the case for a nearly isothermal dark halo of the Galaxy with an
argument independent of previous Galactic models.  It also makes an
unique addition to the handful of estimators for the mass of the
Galaxy (Fich \& Tremaine 1991).  Statistical approaches 
both rely on a large sample and make assumptions about dynamical
equilibrium and velocity distributions for the ensemble 
(e.g. halo satellites or local escaping stars).  The Local
Group timing method depends on the Hubble constant, and like
the Magellanic stream fitting method lacks sensitivity to
the slope of the $\log r-\log M$ relation.  Our model also confirms
proper motion measurements of Sgr and the LMC.

\onecolumn
\twocolumn

On broader aspects the current model provides a platform to piece
together two ancient galaxies which fell into our halo: the Ancient
Magellanic Galaxy (AMG) of Lynden-Bell (1976, 1982) and the Ancient
Sagittarius Galaxy (ASG).  The two ancient galaxies have been torn
apart by the strong tidal field of the Galaxy and have produced two
grand polar streams of tidal debris: remnants of the ASG along
$l\approx 0^o$ may include the globular clusters M54
($l=5.6^o$,$b=-14.1^o$), Terzan 7 ($3.4^o$, $-20.1^o$), Arp 2
($8.5^o$, $-20.8^o$) and Terzan 8 ($5.8^o$, $-24.6^o$), below the
plane and NGC 2419 ($l=180.4^o$, $b=25.2^o$) and Palomar 15 ($18.9^o$,
$24.3^o$) above the plane, and many newly found carbon stars all along
the polar great circle in the APM survey (Irwin 1998, private
communications); remnants of the AMG along $l\approx 270^o$ may
include the LMC, SMC, Carina ($260.1^o$, $-22.2^o$), Draco ($86.4^o,
34.7^o$), Ursa Minor ($105^o, 44.8^o$), and the Magellanic stream.
Curious enough if Sculptor, Sextans, and perhaps Fornax are also on
polar orbits with radius in the range $30$ to $150$ kpc (Lynden-Bell
\& Lynden-Bell 1995), then these might have interacted with the ASG
and the AMG sometime in the past.  Besides constraining the Galactic
potential and proper motions of these remnants as demonstrated, the
current model can help to reconstruct the merging history and star
formation history of these satellites and globular clusters, building
blocks of the Galactic halo.

The model also provides a testbed for theories which explain the newly
discovered polar ring feature of carbon stars around the LMC disc
(Kunkel et al.  1997) and microlensing events towards the LMC (Alcock
et al. 1997) and the well-known warp of the Galaxy (Burton \& te
Lintel Hekkert 1986) all with stars or gas stirred up by the strong
tidal forces among the LMC-SMC-MW triple system (Zhao 1998a and
references therein; Weinberg 1995).  Particularly promising is a
configuration where some stars belonging to a polar ring or a tidal
tail of the Magellanic Clouds are placed at $D \sim (2-10)$ kpc behind
the LMC disc.  These stars would have a high chance of being
microlensed by the numerous stellar lenses in the LMC disc (Zhao
1998a); the optical depth is boosted from a pure LMC disc self-lensing
with a probability of about $0.1\times \tau_{obs}$ (e.g., Wu 1994;
Sahu 1994; Gould 1995) to about $(1-5)\times \tau_{obs}$ by a factor
about ${2 D \over h} \sim (10-50)$, where $h\sim 0.4$ kpc is the scale
height of the LMC disc and $\tau_{obs} \sim 3\times 10^{-7}$ is the
observed optical depth (Alcock et al. 1997).  The Einstein diameter
crossing time of a typical faint stellar lens (say $0.16M_\odot$) in
the LMC disc is about $50-100$ days, for a lens-source velocity of
$\sqrt{2} \times 70 \kms\sim 100\kms$, where $70\kms$ is the typical
rotational speed of stars in the LMC disc and that of the stars in the
polar ring of Kunkel et al. (1997).  This roughly matches the
durations of the dozen or so observed microlensing events towards the
LMC between $34$ and $127$ days (Alcock et al. 1997).  Significant
microlensing at a rate about $1-5$ events per year per million
background stars (ideally red clump giants with distance modulus
$0.1-0.4$ magnitude fainter than the LMC disc) is expected if there
are enough bright source stars in these background tidal features.

\begin{figure}
\epsfysize=15cm \centerline{\epsfbox{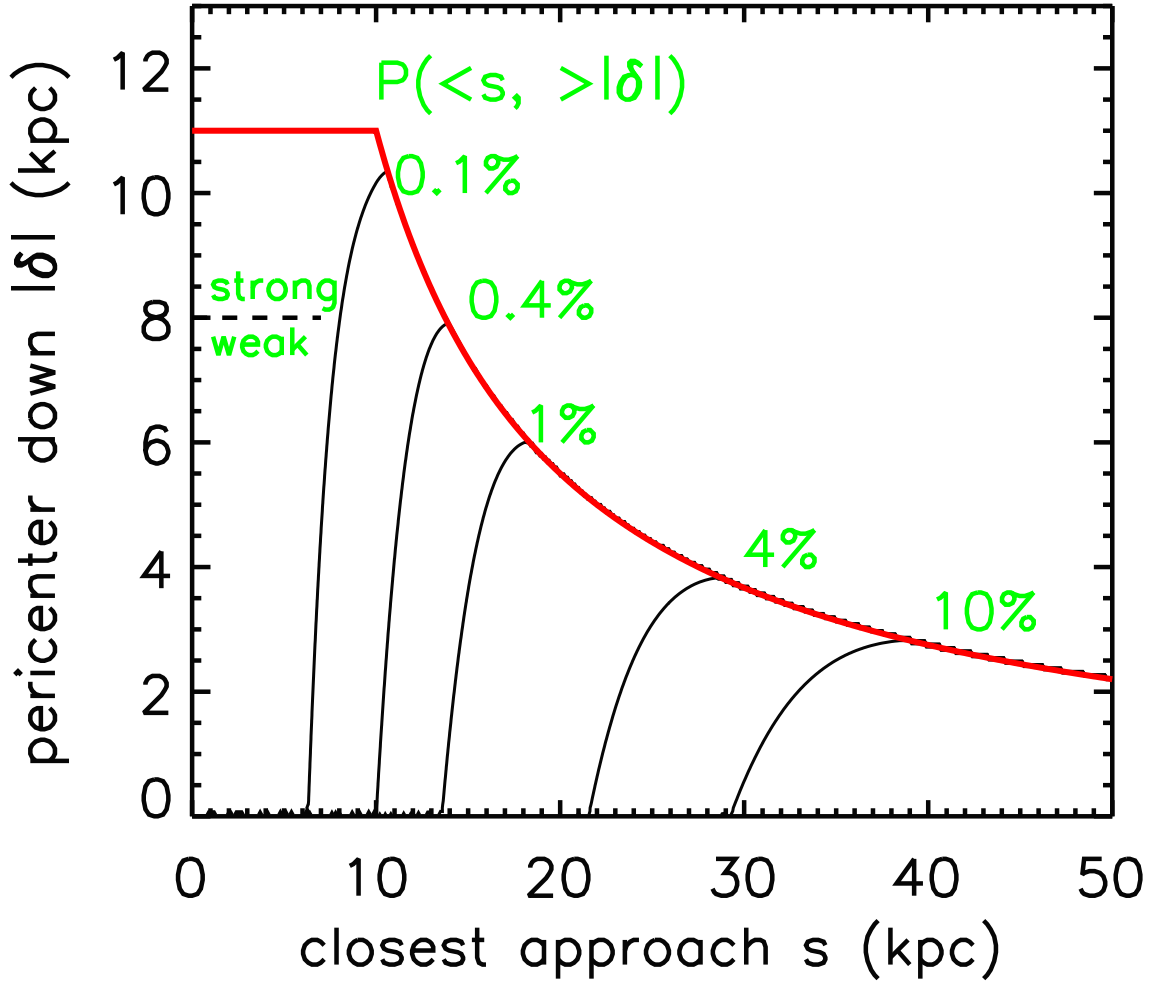}}
\caption{
shows in contours the probability $P(<s,\,>\vert\delta\vert)$ for 
the LMC to come closer than $s$ kpc to Sgr,
and to knock its pericenter down by more than $\vert\delta\vert$ kpc.
The ``shoe-shaped'' boundary,
reminiscent of the rotation curve of the LMC with its tidal radius $a=10$ kpc, 
is set by the scattering power of the LMC
$\propto V^2_{\!c,lmc}(s) \sim (80\kms)^2 
\left[1,{a \over s}\right]_{\rm min}$.
Strong encounters typically involve Sgr coming inside the LMC's tidal radius
($\sim 10$ kpc).
\label{shoe}
}
\end{figure}

It is worth to comment that definite tests of the model require
accurate predictions of the orbital phase within $10^o$ for the past
$3$ Gyrs, equivalent to a proper motion accuracy of $\pm 10$
microarcsec per year.  The current observed proper motions of the LMC
and Sgr have uncertainties of a fraction of one milliarcsec per year,
(the equivalent of about $\pm 180^o$ per Gyr) about two orders of
magnitude too poor to trace back any information of the relative
distance of the LMC and Sgr a few Gyrs ago.  The only information from
these proper motions is that Sgr is somewhere inside a volume $V_0={4
\pi \over 3}R^3$ bound only by its apocenter $R \sim 50$ kpc when the
LMC comes inside the same volume $V_0$.  So the chance of finding Sgr
in a small volume of $V_1={4 \pi \over 3}a^3$ enclosed by the tidal
radius of the LMC ($a\sim 10$ kpc) is ${V_1 \over V_0}=\left({a\over
R}\right)^3\sim 1\%$ (this estimate turns out to be valid within a
factor of two even when taking into account of the fact that Sgr's
orbit is likely to be confined close to a polar plane; the increase of
probability due to a more compact volume $V_0$ is mostly cancelled
when we fold in the smaller chance of finding the LMC's pericenter in
this volume).  So a close encounter is almost equally improbable for
any values of the observed proper motions.  However, it is still
meaningful to access whether the LMC is massive enough to scatter Sgr
to a significantly low orbit with a pericentric change
$\vert\delta\vert \ge 8$ kpc.  The probability for various
combinations of the closest approach ($s$) and the change of pericenter 
($\vert\delta\vert$) is derived in the Appendix, and can
be read out from the contours shown in Figure~\ref{shoe}.

In short we find that the probability of a sudden change of Sgr's
pericenter by between $8$ kpc to $11$ kpc (the maximum) is about
$0.4\%$.  Sgr could have been circulating around the Galaxy at a
``safe'' distance with the pericenter about $20$ kpc before such
strong encounters could bring it down to the present lower orbit with
the pericenter about $10-12$ kpc.  Not only the Galactic pericentric
shocks was a factor $3-4$ weaker on the previous orbit, but also there
was no shocking by the Galactic disc.  These strong encounters
typically involve Sgr coming inside the tidal radius of the LMC with
$s \sim a \sim 10$ kpc, which means that the disintergation of Sgr may
have already started from the tidal shocks of the LMC, which was then
followed by several more shocks by the Galaxy.  These strong
encounters should be contrasted with milder encounters with a change
of the pericenter by between $3$ to $6$ kpc, which has a probability
about 9\% with a typical impact parameter $s \sim 25$ kpc; these can
change the tidal forces by a factor of two.  Fly-bys with $s>30$ kpc
are common, but play no role in explaining the orbit and the survival
of the Sgr.  The fact that there is an upper limit $\vert\delta\vert<
11{\rm kpc}$ means that a drastic change of pericenter by much more
than $10$ kpc would not be possible unless the potential well of the
LMC was in fact much deeper earlier on with a circular rotation speed
$V_{\!c,lmc} \ge 80\kms$.  The same argument also implies that any
deflection by the less massive SMC would be much milder than by the
LMC.

Future astrometric experiments, such as the GAIA, DIVA and SIM
missions as planned by European and American space agencies, which
promise accurate proper motions to a few $\kms$ at $100$ kpc with
space interferometry, will certainly either refine our result
(eq.~\ref{logr-logm}) on the mass and potential of the Galactic halo
or rule out the proposed scenario.  In the nearer future the model is
also observationally testable by mapping out tidal debris along the
great circle of Sgr.  If the disruption of Sgr started from the
collision with the LMC, there should be plenty of time for material to
spread out to a very long tidal tail: the trailing arm of the debris
should be visible at $45^o$ below the Galactic plane, and the leading
arm $15^o$ above if the N-body simulation of Vel\'azquez \& White
(1995) is simply rescaled from $1$ Gyr to $2-3$ Gyrs.

We thank Dave Syer, Hector Vel\'azquez, Kathryn Johnston, Butler
Burton, Gerry Gilmore and Mike Irwin for stimulating
discussions at various stages of the preparation, Amina Helmi, Ronnie
Hoogerwerf and Maartje Sevenster for being the most helpful colleagues 
sharing the same floor, David Spergel for both encouragements and
critical remarks, Tim de Zeeuw for critical readings 
of the manuscript, and the referee Rodrigo Ibata for
careful scrutiny on the probability of the collision.

\onecolumn

\appendix
\section{Strength and probability of encounter estimated with impulse approximation}

For a simple logarithmic potential well of the LMC with
a circular speed $V_{\!c,lmc}$, the impulse approximation predicts a
kick velocity $\Delta V = {\pi V^2_{\!c,lmc} \over V_{rel}}$
(cf. Binney \& Tremaine 1987), where
the relative speed of encounter $V_{rel} \sim 300 \kms$, much faster
than the internal motions of stars in the LMC and Sgr.  
A kick at distance $R$ makes a sudden change of
the angular momentum of Sgr by an amount as much as $R \Delta V$,
equivalent to a sudden change of pericenter by an amount
$\sim {\Delta V \over V_{\!c,mw}} R$, if the kick is oriented to the
right direction.  However two effects diminish the change: 
the rotation curve of the LMC falls as ${a \over s} V^2_{\!c,lmc}$ 
outside the tidal radius $a$ of the LMC, and the kick is in random directions.
In general the pericenter changes by
\beq\label{deltamax}
\delta = \delta_{max} \left[1,{a \over s}\right]_{\rm min} \cos \alpha,~
\delta_{max} \equiv {2fV^2_{\!c,lmc} \over V_{\!c,mw} V_{rel}} R 
\sim 11 {\rm kpc},
\eeq
where $\alpha$ is an angle roughly evenly distributed between $0$ and $\pi$.
$\delta$ is positive for $0\le \alpha < {\pi \over 2}$, 
which happens $50\%$ of the time,
corresponding to the less interesting situations where Sgr was dragged up.
In estimating $\delta_{max}$
we set  the fudge factor $f=1$ (the value for Keplerian potential, 
${\pi \over 2}$ for logarithmic potential) 
everywhere to be conservative, but
adopt a slightly high value for the peak circular speed with 
$V_{\!c,lmc} \sim 40\kms \sin^{-1}\!i \sim 80\kms$ 
for a disc inclination ($i$) of about $33^o$.
So the compounded probability, $dP_s(>\vert\delta\vert)$, 
for the LMC to come to a distance between $s$ to $s+ds$ of Sgr, 
and to knock Sgr's pericenter down by an amount greater than 
$\vert\delta\vert$ should be given by
\beq
dP_s(>\vert\delta\vert) = 
{\arccos \epsilon \over \pi} d \left({s \over R}\right)^3,~
\epsilon(s) \equiv \left[{\vert\delta\vert \over \delta_{max}}, {\vert\delta\vert s \over \delta_{max} a} \right]_{max},
\eeq
where ${\arccos \epsilon \over \pi}$ takes into account of the
random chance that a kick is in the required range of directions
as specified by $\pi - \arccos \epsilon(s) \le \alpha \le \pi$.
Integrating over all possible impact parameter $s$ we find that the probability
to pull down the pericenter by more than $\vert\delta\vert$
\beq
P(>\vert\delta\vert) = \left({a \over R}\right)^3 Q\left({\delta^2_{max} \over \delta^2}\right),~Q(\gamma)\equiv
{(2\gamma+1) \over 3 \pi} (\gamma-1)^{1\over 2},
\eeq
where $Q$ is about unity for strong encounters and increases
as $\propto \delta^{-3}$ for weak encounters since these can happen for bigger
$s$ which prescribes a bigger volume.  

{}

\label{lastpage}
\end{document}